\documentstyle[12pt]{article}

\begin{document}
\baselineskip=18pt plus 0.2pt minus 0.1pt
\parskip = 6pt plus 2pt minus 1pt
\newcommand{\reseteqnum}{\setcounter{equation}{0}}
\newcommand{\bx}{\mbox{\boldmath $x$}}
\newcommand{\bX}{\mbox{\boldmath $X$}}
\newcommand{\bY}{\mbox{\boldmath $Y$}}
\newcommand{\rhovev}{(\rho v)^2}
\newcommand{\mr}{\frac{g^2}{\lambda}}
\newcommand{\bra}[1]{\left\langle #1 \right|}
\newcommand{\ket}[1]{\left| #1 \right\rangle}
\newcommand{\VEV}[1]{\left\langle #1 \right\rangle}
\newcommand{\braket}[2]{\VEV{#1 | #2}}
\newcommand{\calP}{{\cal P}}
\def\Dslash{\,{\raise.15ex\hbox{/}\mkern-12mu D}}
\def\Dbarslash{\,{\raise.15ex\hbox{/}\mkern-12mu {\bar{D}}}}
\def\slashpartial{\,{\raise.15ex\hbox{/}\mkern-10.5mu \partial}}

\begin{titlepage}
\title{
\hfill
\parbox{4cm}{\normalsize KUCP-0100\\KUNS-1402\\HE(TH)96/05 \\hep-th/9607076}\\
\vspace{1cm}
Multi-instanton calculus \\ in N=2 supersymmetric QCD
}
\author{Hideaki Aoyama\thanks{e-mail address:
\tt aoyama@phys.h.kyoto-u.ac.jp}
\\
{\normalsize\sl Department of Fundamental Sciences,}
\\
{\normalsize\sl Faculty of Integrated Human Studies,
Kyoto University, Kyoto 606-01, Japan}
\vspace{0.3cm}
\\
Toshiyuki Harano\thanks{e-mail address: \tt harano@gauge.scphys.kyoto-u.ac.jp}
\\
{\normalsize\sl Department of Physics, Kyoto University, Kyoto 606-01, Japan}
\vspace{0.3cm}
\\
Masatoshi Sato\thanks{e-mail address: \tt msato@gauge.scphys.kyoto-u.ac.jp}
\\
{\normalsize\sl Department of Physics, Kyoto University, Kyoto 606-01, Japan}
\vspace{0.3cm}
\\
Shinya Wada\thanks{e-mail address: \tt shinya@phys.h.kyoto-u.ac.jp}
\\
{\normalsize\sl Graduate School of Human and Environmental Studies,
Kyoto University}\\
\normalsize\sl Kyoto 606-01, Japan}
\date{\normalsize July 1996}
\maketitle
\thispagestyle{empty}
\begin{abstract}
Microscopic tests of the exact results are performed in N=2 SU(2)
supersymmetric QCD. 
We construct the multi-instanton solution in N=2 supersymmetric QCD and
calculate the two-instanton contribution ${\cal F}_2$ to the prepotential 
${\cal F}$ explicitly. 
For $N_f=1,2$, instanton calculus agrees with the prediction of the
exact results, however, for $N_f=3$, we find a discrepancy between them.  
\end{abstract}

\end{titlepage}

\newpage
\reseteqnum

\section{Introduction}

Recently, much progress has been made in the study of the strongly coupled
supersymmetric gauge theories.
Under the holomorphy and the
duality, the low energy effective actions of N=2 supersymmetric Yang-Mills
theory and supersymmetric QCD in Coulomb phase are determined exactly
for SU(2) gauge group\cite{SW} and later for larger gauge
groups\cite{KLYT}-\cite{AS}. 
These low energy effective theories reveal the interesting results
like the monopole condensation\cite{SW} and new supersymmetric conformal field
theories\cite{APSW,EHIY} and so on. 

The exact results predict the non-perturbative corrections from
instanton.
Furthermore, it is known that the instanton calculus in the supersymmetric
theories is fully controllable when the theories is weakly
coupled \cite{ADS,NSVZ,AKMRV}. 
Therefore, the instanton calculus gives a non-trivial test of the
exact results. 
Until now, the instanton calculi were performed in the pure Yang-Mills
theories and all the microscopic calculi agree 
with the exact results \cite{FP,IS,DKM,FT}.

In this letter, we examine the consistencies between the instanton calculus and
the exact results of N=2 supersymmetric QCD. 
Especially, we focus on the N=2 supersymmetric SU(2) QCD.
In N=2 supersymmetric SU(2) QCD, there is a parity symmetry
between hypermultiplets, then only contributions from even number of
instanton exist \cite{SW}. 
Thus the instanton corrections start from the two-instanton sector.
In the following, we perform the two-instanton calculus in N=2
supersymmetric SU(2) QCD for $N_f\leq 3$ flavors\footnote{In the following, we
give the supersymmetric instanton in the two-instanton sector. 
However, the construction is applicable to
  the arbitrary number of multi-instantons. More details
  are given in \cite{AHSW}.}
and compare it with the exact results.

\section{The construction of multi-instanton}
First we will briefly summarize N=2 vector multiplet of
supersymmetric instanton\cite{DKM}.
The defining equations of N=2 vector multiplet of supersymmetric
instanton are following; 
\begin{eqnarray}
  \label{eqofmotionA}
&&\hspace{5ex}F_{\mu\nu}=-\tilde{F}_{\mu\nu},\\ 
  \label{eqofmotionlam}
&&\hspace{1ex}\Dbarslash\lambda=0,\quad\Dbarslash\psi=0,\\
  \label{eqofmotionphi}
&&D^2\phi-\sqrt{2}i[\lambda,\psi]=0.
\end{eqnarray}
When the coupling constant is small enough, the solution of the above
equation dominates the path integral. 
In supersymmetric theories, the coupling dependence of the instanton
contribution is fixed, then it is enough to consider the small
coupling case. 
The first three equations in the above mean that $A_{\mu}$ is
an instanton and $\lambda$ and $\psi$ are adjoint fermion zero modes.  
The last one is the supersymmetric version of the 't Hooft
equation\cite{tHooft}.
The multi-instanton solution is constructed by ADHM method\cite{ADHM,CWS}.
In the two-instanton sector, its explicit form is the following;
\begin{eqnarray}
  \label{ADHM}
&&A_{\mu}=iN^{\dagger\dot{r}r}\partial_{\mu}N_{r\dot{s}},
\end{eqnarray}
where $N$ is a quaternionic
3-dimensional column vector
\footnote{We represent a quaternion $N$ as 2$\times$2 matrix
  $N_{s\dot{s}}=-iN_{\mu}\sigma_{\mu s \dot{s}}$}
 obeying 
\begin{eqnarray}
N^{\dagger}M=0,\quad N^{\dagger}N=1.
\end{eqnarray}
Here $M$ is a 3$\times$2 matrix made up of quaternions;
\begin{eqnarray}
\label{M}
M&=&\left(
  \begin{array}{cc}
    \omega_1&\omega_2\\
    x_0-x+a_3&a_1\\
    a_1&x_0-x-a_3
  \end{array}
\right),\\
 \label{a1}
  a_1&=&\frac{a_3}{4|a_3|^2}
\left(\bar{\omega}_2\omega_1-\bar{\omega}_1\omega_2\right).
\nonumber
\end{eqnarray}
The relation between $a_1$ and $a_3$ is required by the reality
condition of $R=M^{\dagger}M$ and this ensures the anti-self-duality of
$F_{\mu\nu}$. 
The adjoint fermionic zero modes are the following\cite{CGT};
\begin{eqnarray}
  \label{ADHMlam}
&&  \lambda^{\dot{r}}_{\alpha\dot{s}}=N^{\dagger\dot{r}r}\left\{{\cal
    M}_r R^{-1}C^T\delta_{\alpha}^s + \epsilon_{r \alpha}CR^{-1}({\cal
    M}^T)^s\right\}N_{s\dot{s}}, \\
  \label{ADHMpsi}
&&\psi^{\dot{r}}_{\alpha\dot{s}}=N^{\dagger\dot{r}r}\left\{{\cal
    N}_r R^{-1}C^T\delta_{\alpha}^s + \epsilon_{r \alpha}CR^{-1}({\cal
    N}^T)^s\right\}N_{s\dot{s}}, 
\end{eqnarray}
where
\begin{eqnarray}
  \label{collect}
{\cal M}_s=\left(
  \begin{array}{cc}
    \mu_{1s}&\mu_{2s}\\
    4\xi_s+m_{3s}&m_{1s}\\
    m_{1s}&4\xi_s-m_{3s}
  \end{array}
\right),
\quad {\cal N}_s=\left(
  \begin{array}{cc}
    \nu_{1s}&\nu_{2s}\\
    4\xi'_s+n_{3s}&n_{1s}\\
    n_{1s}&4\xi'_s-n_{3s}
  \end{array}
\right),
\end{eqnarray}
\begin{eqnarray}
   \label{m1}
&&m_1=\frac{a_3}{2|a_3|^2}
\left(2\bar{a}_1m_3+\bar{\omega}_2\mu_1-\bar{\omega}_1\mu_2\right), \\
  \label{n1}
&&n_1=\frac{a_3}{2|a_3|^2}
\left(2\bar{a}_1n_3+\bar{\omega}_2\nu_1-\bar{\omega}_1\nu_2\right), \\
&&C=\left(
  \begin{array}{cc}
    0&0\\
    1&0\\
    0&1
  \end{array}
\right).
\end{eqnarray}
The solution of Eq.(\ref{eqofmotionphi}) is a sum of the solution of the 
homogeneous equation $D^2\phi_0=0$ and a particular solution $\phi_f$;
\begin{equation}
\phi=\phi_0+\phi_f.  
\end{equation}
The explicit forms of $\phi_0$ and $\phi_f$ are 
\begin{eqnarray}
&&\phi_0=-iN^{\dagger\dot{r}r}A_r^s N_{s\dot{s}},\\
&&\phi_f=\frac{\sqrt{2}i}{4}N^{\dagger\dot{r}r}\left\{{\cal
    N}_rR^{-1}({\cal M}^T)^s-{\cal
    M}_rR^{-1}({\cal N}^T)^s+i F\delta_r^s\right\}N_{s\dot{s}},
\end{eqnarray}
where
\begin{eqnarray}
&&A_r^s=\left(
\begin{array}{ccc}
A_{00\, r}^{\quad s}&0&0\\
0&0&\gamma\delta_r^s\\
0&-\gamma\delta_r^s&0
\end{array}
\right),
\quad F=\left(
\begin{array}{ccc}
0&0&0\\
0&0&\alpha\\
0&-\alpha&0
\end{array}
\right),\\
&&A_{00}=i\VEV{\phi},\quad \gamma=-\frac{\omega}{H},\\
&&\alpha=-\frac{i}{H}\left(\mu_1\nu_2-\mu_2\nu_1+2m_3n_1-2m_1n_3\right),\\
  \label{HL}
&&L=|\omega_1|^2+|\omega_2|^2,\quad
H=L+4|a_1|^2+4|a_3|^2,\\
\nonumber
&&\Omega=\omega_1\bar{\omega}_2-\omega_2\bar{\omega}_1,\quad
\omega=\frac{1}{2}{\rm tr}\left(\Omega A_{00}\right).
\end{eqnarray}

In N=2 supersymmetric QCD, there appear $N_f$ hypermultiplets in the theory. 
The N=2 hypermultiplets of supersymmetric instanton are characterized by the
following equations;
\begin{eqnarray}
  \label{eqofmotionq}
&&\Dbarslash q=0,\hspace{15ex}\Dbarslash \tilde{q}=0,\\
  \label{eqofmotionQ}
&&D^2Q-\sqrt{2}i\lambda q=0,\hspace{3ex}
D^2\tilde{Q}+\sqrt{2}i\tilde{q}\lambda=0,\\
  \label{eqofmotionQdagger}
&&D^2Q^{\dagger}-\sqrt{2}i\tilde{q}\psi=0,\quad
D^2\tilde{Q}^{\dagger}-\sqrt{2}i\psi q=0.
\end{eqnarray}
The first two equations indicate that $q$ and $\tilde{q}$ are the
fundamental fermionic zero modes\cite{OCFGT};
\begin{eqnarray}
  \label{SQCDsol}
q_{f\alpha} ^{\dot{r}}=\Psi_{\alpha}^{\dot{r}}\zeta_f,\quad
\tilde{q}^{\alpha}_{f\dot{r}}
=-\epsilon^{\alpha\beta}\epsilon_{\dot{r}\dot{s}}
\Psi_{\beta}^{\dot{s}}\tilde{\zeta}_f,
\end{eqnarray}
where the indices $\alpha$, $\dot{r}$ and $f$ of $q$ and $\tilde{q}$
are a spinor, color and flavor index respectively.
$\Psi$ is a following normalized function.  
\begin{eqnarray}
\label{zeromode}
\Psi_{\alpha}^{\dot{r}}=-\frac{1}{\pi}N^{\dagger \dot{r}r}\epsilon_{r
  \alpha}CR^{-1},\quad
\int d^4 x\epsilon^{\alpha\beta}\epsilon_{\dot{r}\dot{s}}
   \Psi_{\alpha k}^{\dot{r}}\Psi_{\beta l}^{\dot{s}}
=-\delta_{k l}.
\end{eqnarray}
The solution of Eq.(\ref{eqofmotionQ}) and (\ref{eqofmotionQdagger}) is
given by
\begin{eqnarray}
&&Q^{\dot{r}}_f
=\frac{\sqrt{2}i}{4\pi}N^{\dagger \dot{r}r}{\cal M}_rR^{-1}\zeta_f,
\quad\tilde{Q}_{f\dot{r}}
=-\frac{\sqrt{2}i}{4\pi}\epsilon_{\dot{r}\dot{s}}
N^{\dagger \dot{s}r}{\cal M}_rR^{-1}\tilde{\zeta}_f,\\
&&Q^{\dagger}_{f\dot{r}}
=\frac{\sqrt{2}i}{4\pi}\epsilon_{\dot{r}\dot{s}}
N^{\dagger \dot{s}r}{\cal N}_rR^{-1}\tilde{\zeta}_f,
\quad\tilde{Q}^{\dagger\dot{r}}_f=\frac{\sqrt{2}i}{4\pi}
N^{\dagger \dot{r}r}{\cal N}_rR^{-1}\zeta_f.
\end{eqnarray}
In N=2 supersymmetric QCD, the anti-scalar component of N=2 vector
multiplet satisfies the
following equation.
\begin{eqnarray}
\label{eqofmotionphidagger}
&&D^2\phi^{\dagger a}-\sqrt{2}i\tilde{q}T^a q=0.
\end{eqnarray}
The solution of Eq.(\ref{eqofmotionphidagger}) is given by,
\begin{eqnarray}
&&\phi^{\dagger}=\phi_0^{\dagger}+\phi_q^{\dagger},\\
&&\phi^{\dagger}_q=-iN^{\dagger \dot{r}r}PN_{r \dot{s}},
\end{eqnarray}
where
\begin{eqnarray}
&&P=\left(
\begin{array}{ccc}
0&0&0\\
0&0&\beta\\
0&-\beta&0
\end{array}
\right),
\quad\beta=\frac{\sqrt{2}}{16}\frac{\tilde{\zeta}_f\zeta_f}{H}.
\end{eqnarray}
The part of Lagrangian which gives the important contribution is the
following;  
\begin{eqnarray}
\nonumber
&&g^2{\cal L}_m={\rm tr}\left\{2(D_{\mu}\phi)^{\dagger}D_{\mu}\phi
    +2\sqrt{2}ig\lambda[\psi,\phi^{\dagger}]\right\}
+(D_{\mu}Q)^{\dagger}D_{\mu}Q
+D_{\mu} \tilde{Q}(D_{\mu}\tilde{Q})^{\dagger}\\ 
\nonumber
&&\hspace{5ex}+\sqrt{2}i\left(\tilde{q}\phi q+Q^{\dagger}\lambda q
-\tilde{q}\lambda\tilde{Q}^{\dagger}
+\tilde{q}\psi Q+\tilde{Q}\psi q\right)\\
&&\hspace{4ex}=\partial_{\mu}\left\{{\rm tr}(2\phi^{\dagger}D_{\mu}\phi)
+(D_{\mu}Q)^{\dagger}Q+(D_{\mu}\tilde{Q})^{\dagger}\tilde{Q}\right\}
+\sqrt{2}i\left(\tilde{q}\phi q+Q^{\dagger}\lambda q
+\tilde{Q}\psi q\right).
  \label{lagmatter}
\end{eqnarray}
The last equality follows from an integration by parts and the
equation of supersymmetric instanton.
To integrate the last term, we use the auxiliary solution $\bar{q}$,
\begin{eqnarray}
  \label{qbar}
  \bar{q}^{\dot{\alpha}}_{\dot{r}}=\frac{1}{4\pi}\epsilon_{\dot{r}\dot{s}}
    N^{\dagger\dot{s}r}\left\{
    {\cal N}_rR^{-1}({\cal M}^T)^s-{\cal M}_rR^{-1}({\cal N}^T)^s
    +i  F\delta_r^s\right\}M_{s\dot{t}}
    \epsilon^{\dot{t}\dot{\alpha}}R^{-1}\tilde{\zeta},
\end{eqnarray}
which satisfies the equation;
\begin{eqnarray}
\label{eqqbar}
\Dslash \bar{q}
+\sqrt{2}Q^{\dagger}\lambda+\sqrt{2}\tilde{q}\phi_f
+\sqrt{2}\tilde{Q}\psi=-\Psi\eta,
\end{eqnarray}
where
\begin{eqnarray}
\hspace{5ex}\eta=\frac{1}{2}\alpha
\left(
  \begin{array}{cc}
    0&1\\
    -1&0
  \end{array}
\right)\tilde{\zeta}.
\end{eqnarray}
Using the auxiliary solution $\bar{q}$ and $\phi^{\dagger}_q$, the last term
of Eq.(\ref{lagmatter}) becomes 
\begin{equation}
\partial_{\mu}{\rm tr}\left\{2(D_{\mu}\phi_q^{\dagger})\phi_0\right\}
-i\slashpartial(\bar{q}q)
-i\epsilon^{\alpha\beta}\epsilon_{\dot{r}\dot{s}}
\Psi^{\dot{s}}_{\beta}\eta q_{\alpha}^{\dot{r}} .
\end{equation}
{}From the normalization condition of $\Psi$ and the asymptotic
behaviors of $\bar{q}$ and the supersymmetric instanton, the action of 
supersymmetric instanton becomes
\begin{eqnarray}
  \label{action}
&&g^2 S=16\pi^2+S_{higgs}+S_{yukawa},\nonumber\\
&&S_{higgs}=16\pi^2\left(L|A_{00}|^2-\frac{\omega^2}{H}\right),\\
&&S_{yukawa}=-4\sqrt{2}\pi^2\left\{\nu_kA_{00}\mu_k
+\frac{\omega}{H}\left(\mu_1\nu_2-\mu_2\nu_1+2m_3n_1-2m_1n_3\right)\right\}
\nonumber\\
&&\hspace{8ex}+\frac{1}{2H}\left(
\mu_1\nu_2-\mu_2\nu_1+2m_3n_1-2m_1n_3\right) 
\tilde{\zeta}_f\zeta_f
+\sqrt{2}\frac{\omega}{H}\tilde{\zeta}_f\zeta_f.
\nonumber
\end{eqnarray}
Comparing to the pure Yang-Mills case, the last two term in
$S_{yukawa}$ are added.
Note that a biquadratic term in Grassmannian variables appears in the action.
This is a new feature in the N=2 supersymmetric QCD.
The measure of the collective coordinate is given by\cite{Osborn,DKM},
\begin{eqnarray}
  \label{measure}
&&C_J\int d^4x_0d^4a_3 d^4\omega_1
d^4\omega_2d^2\xi d^2m_3d^2\mu_1d^2\mu_2d^2\xi'd^2n_3d^2\nu_1d^2\nu_2\\
\nonumber
&&\hspace{5ex}\times 
\prod_{f=1}^{N_f}d^2\zeta_f d^2\tilde{\zeta}_f
\frac{\left|\left|a_3\right|^2-\left|a_1\right|^2\right|}{H}
\exp\left(-S_{higgs}-S_{yukawa}\right),
\end{eqnarray}
where the coupling constant $g$ is absorbed by the redefinition of the 
collective coordinates and
\begin{eqnarray}
  \label{const}
  C_J=2^{6+2 N_f}\pi^{-8}\Lambda_{N_f}^{8-2N_f}.
\end{eqnarray}

\section{Instanton calculus}
We calculate two-instanton contribution to $\VEV{u}=\VEV{{\rm tr}  \phi^2}$. 
Taking into account the super transformation, it is easy to find that
the adjoint scalar $\phi$ contains the following part; 
\begin{eqnarray}
  \label{superphi}
  \phi=-\sqrt{2}i\xi\psi+\cdots
=\sqrt{2}i\xi\bar{\sigma}_{\mu\nu}\xi'F_{\mu\nu}+\cdots,
\end{eqnarray}
where $\cdots$ includes the other fermionic zero modes and $\phi_0$.
The normalization of supersymmetric modes, $\xi$ and $\xi'$ 
is determined by the Eq(\ref{collect}).
Then, $u$ is given by,
\begin{eqnarray}
&&  u=-2{\rm
    tr}\left[\left(\xi\bar{\sigma}_{\mu\nu}\xi'F_{\mu\nu}\right)^2\right]
+\cdots\\
&&\hspace{2ex}=-\xi^2\xi'^2{\rm
tr}\left(F_{\mu\nu}F_{\mu\nu}\right)+\cdots\, .
\end{eqnarray}
Therefore supersymmetric zero modes are saturated by inserting $u$, and 
we obtain the following result by performing the integration over the center
of the instanton; 
\begin{eqnarray}
  \label{ssmode}
  \int d^4 x_0\int d^2\xi d^2\xi'u(x)=-\int d^4 x_0 {\rm
    tr}\left[F_{\mu\nu}(x-x_0)F_{\mu\nu}(x-x_0)\right] =-32\pi^2 \,
.
\end{eqnarray}
The other fermionic modes are lifted by the Yukawa terms in the
action, and integrating out those modes except $\zeta_f$,
$\tilde{\zeta}_f$, we obtain
\begin{eqnarray}
  \nonumber
&&\int d^2 m_3 d^2 \mu_1 d^2\mu_2 d^2 n_3 d^2 \nu_1 d^2\nu_2
  \exp\left(-S_{yukawa}\right) \\
\label{yukawa}
&&\hspace{5ex}=-\left(\frac{16\sqrt{2}\pi^6}{|a_3|^2H|\Omega|}\right)^2 
f(y)\exp\left(-\sqrt{2}\frac{\omega}{H}\tilde{\zeta}_f\zeta_f\right),
\end{eqnarray}
where
\begin{eqnarray}
  \label{f(y)}
 &&f(y)=\omega^2y^2 
\left\{
  \left(|\Omega|^2|A_{00}|^2+\frac{L\omega^2 y}{H}\right)^2
  +\frac{L^2-|\Omega|^2}{H^2} \omega^2 y^2 
\left(|A_{00}|^2|\Omega|^2-\omega^2\right)
\right\},\\
  \label{y}
 &&\hspace{5ex}y=1-\frac{\sqrt{2}}{16\pi^2\omega}
\tilde{\zeta}_f\zeta_f.
\end{eqnarray}
The remaining Grassmann integrations are performed in the following;
\begin{eqnarray}
  \nonumber
&& \int \prod_{f=1}^{N_f}d^2\tilde{\zeta}_fd^2\zeta_f
  f(y)\exp\left(-\sqrt{2}\frac{\omega}{H}\tilde{\zeta}_f\zeta_f\right)
=\left(-\frac{1}{2}
\frac{\omega^2}{H^2}\right)^{N_f}\sum_{k=0}^{2N_f}\,_{2N_f}\hspace{-0.3ex}C_k 
\left(\frac{H}{16\pi^2\omega^2}\right)^k\left.\frac{\partial^k f}
{\partial y^k}\right|_{y=1}.\\
\label{zetaint}
\end{eqnarray}
We change the integration variables from $a_3,\omega_1,\omega_2$ to
$H,L,\Omega$, and then the measure of the integral becomes,
\begin{eqnarray}
  \label{bosonicmeasure}
&&\int d^4 a_3\frac{\left|
      \left|a_3\right|^2-\left|a_1\right|^2\right|}{|a_3|^4} 
=\frac{\pi^2}{2}\int_{L+2|\Omega|}^{\infty}dH,\\
&&\int d^4\omega_1
d^4\omega_2=\frac{\pi^3}{8}\int_0^{\infty}dL\int_{|\Omega|\leq
L}d^3\Omega .
\end{eqnarray}
With the change to a polar
coordinate:$\omega=|\Omega||A_{00}|\cos\theta$ and 
the rescaling:$\Omega'=\Omega /L$ and $H'=H/L$,
the measure is given by,
\begin{eqnarray}
\nonumber
&&\frac{\pi^5}{16}\int_0^{\infty}d L\int_{|\Omega|\leq L}d^3\Omega
  \int_{L+2|\Omega|}^{\infty}dH\\
\label{mrescale}
&&\hspace{3ex}=\frac{\pi^6}{8}\int_0^{\infty}dL L^4\int_{-1}^1d(\cos
\theta)\int_0^1|\Omega'|^2d|\Omega'|\int_{1+2|\Omega'|}^{\infty}dH' ,
\end{eqnarray}
and  $f(y)$ becomes
\begin{eqnarray}
  \label{frescale}
  f(y)=|A_{00}|^6|\Omega'|^6L^6\cos^2\theta \, G(y;|\Omega'|,H',\theta),
\end{eqnarray}
where
\begin{eqnarray}
  \label{G(y)}
  G(y;|\Omega'|,H',\theta)=y^2\left\{\left(1+\frac{y}{H'}\cos^2\theta\right)^2
+\frac{1-|\Omega'|^2}{4H'^2}y^2\sin^2 2\theta\right\}.
\end{eqnarray}
Using Eq.(\ref{measure}), (\ref{ssmode}), (\ref{yukawa}), (\ref{zetaint}),
(\ref{mrescale}) and (\ref{frescale}) and performing the integration
of $L$ ,  we obtain the two-instanton
correction to $\VEV{u}$,
\begin{eqnarray}
  \label{vevu}
  \VEV{u}_{2}=\frac{1}{2}a^2\left(\frac{\Lambda_{N_f}}{ a}\right)^{8-2N_f}
  \cdot\left(-\frac{1}{2}\right)^{N_f} I(N_f) ,
\end{eqnarray}
where $I(N_f)$ is defined by
\begin{eqnarray}
  \label{inf}
&&I(N_f)=\int_{-1}^1d(\cos\theta)\cos^2\theta\int_0^1d|\Omega'||\Omega'|^6
\int_{1+2|\Omega'|}^{\infty}\frac{dH'}{H'^3} \left(\frac{|\Omega'|\cos \theta
    }{H'}\right)^{2N_f}\sum_{k=0}^{K}\,_{2N_f}\hspace{-0.3ex}C_k\,
\\
\nonumber
&&\hspace{10ex}\times(5-k)!
\left(1-\frac{|\Omega'|^2\cos^2\theta}{H'}\right)^{k-6}
\left(\frac{H'}{|\Omega'|^2
\cos^2\theta}\right)^k\left.\frac{\partial^k}{\partial  y^k} 
G(y;|\Omega'|,H',\theta)\right|_{y=1},
\end{eqnarray}
and $K={\rm min}[4,2N_f]$. The integral $I(N_f)$ is complicated but
elementary.
Finally we obtain 
\begin{eqnarray}
\label{instcal}
  \VEV{u}_2=\frac{1}{2}a^2 \times \left\{
    \begin{array}{ccc}
\vspace{1ex}     
      \displaystyle
\frac{5}{2}\left(\frac{\Lambda_0}{a}\right)^8&{\rm for}&N_f=0\,,\\
\vspace{1ex}
      \displaystyle 
-\frac{3}{4}\left(\frac{\Lambda_1}{a}\right)^6&{\rm for}&N_f=1\,,\\
\vspace{1ex}
      \displaystyle 
\frac{1}{8}\left(\frac{\Lambda_2}{a}\right)^4&{\rm for}&N_f=2\,,\\
      \displaystyle 
-\frac{5}{2^{4} 3^{3}}\left(\frac{\Lambda_3}{a}\right)^2&{\rm for}& N_f=3.
    \end{array}\right. 
\end{eqnarray}

\section{Exact results versus instanton calculus}
The low energy effective Lagrangians for $N=2$ supersymmetric gauge
theories are determined by the holomorphic function ${\cal F}$, which is
called the prepotential.
According to \cite{SW}, the prepotential ${\cal F}$ are 
determined by the elliptic curves;
\begin{eqnarray}
  \label{ellipticSYM}
 N_f=0&:& y^2=x^2(x-u)+\frac{1}{4}\tilde{\Lambda}_0^4 x,\\
  \label{ellipticSQCD}
 N_f=1,2,3&:& y^2=x^2(x-u)-\frac{1}{64}
\tilde{\Lambda}_{N_f}^{2(4-N_f)}(x-u)^{N_f-1},
\end{eqnarray}
in the $SU(2)$ gauge theories.
In the semiclassical limit, prepotential $\cal F$ is expanded by the
one-loop correction and $k$-instanton contributions;
\begin{eqnarray}
  \label{prepot}
  {\cal F}(a)=\frac{i
    a^2}{4\pi}\left\{(4-N_f)
\ln\left(\frac{a^2}{\tilde{\Lambda}_{N_f}^2}\right)
+\sum_{k=0}^{\infty}{\cal F}_k(N_f)
\left(\frac{\tilde{\Lambda}_{N_f}}{a}\right)^{(4-N_f)k}\right\}.
\end{eqnarray}
In this convention, the coefficients ${\cal F}_{2n+1}$ vanish for
$N_f\neq 0$. The vacuum expectation value of $u$ is given as the
function of $a$\cite{Matone} by, 
\begin{eqnarray}
  \label{u(a)}
  u(a)&=&\frac{8\pi i}{4-N_f}\left({\cal F}(a)-\frac{1}{2}a
    \partial_a{\cal F}(a)\right) \\
\nonumber
&=&2 a^2\left\{1-\frac{1}{2}\sum_{k=1}^{\infty}k{\cal F}_k(N_f)
\left(\frac{\tilde{\Lambda}_{N_f}}{a}\right)^{(4-N_f)k}\right\}.
\end{eqnarray}
{}From the Picard-Fuchs equation, we can obtain ${\cal F}_k$
recursively\cite{IY}. The Picard-Fuchs equation is given by,
\begin{eqnarray}
  \label{picard}
  p(u)\partial_a^2 u-a(\partial_a u)^3=0,
\end{eqnarray}
where
\begin{eqnarray}
  \label{p(u)}p(u)=\left\{
    \begin{array}{lll}
\displaystyle 4(u^2-\tilde{\Lambda}_0^4)&{\rm for}&N_f=0\,,\\
\vspace{1ex}
\displaystyle 4 u^2+\frac{27 \tilde{\Lambda}^6_1}{64 u}&{\rm for}&N_f=1\,,\\
\vspace{1ex}
\displaystyle 4(u^2-\frac{\tilde{\Lambda}_2^4}{64})&{\rm for}&N_f=2\,,\\
\vspace{1ex}
\displaystyle u(4u-\frac{\tilde{\Lambda}_3^2}{64})&{\rm for}&N_f=3 \, .
  \end{array}\right.
\end{eqnarray}
Only when $N_f$=0, one-instanton contribution ${\cal
  F}_1$ does not vanish, and this coefficient agrees with
microscopic one-instanton calculus, if we identify the dynamical scale 
$\tilde{\Lambda}_0=\sqrt{2}\Lambda_0$\cite{FP}.
The two-instanton correction to $\VEV{u}$ is given by ${\cal F}_2$,
and we obtain 
\begin{eqnarray}
\label{exactresult}
  \VEV{u}_2=2 a^2 \times \left\{
    \begin{array}{ccc}
\vspace{1ex}     
      \displaystyle
5\cdot2^{-13}\left(\frac{\tilde{\Lambda}_0}{a}\right)^8&{\rm for}&N_f=0\,,\\
\vspace{1ex}
      \displaystyle 
-3\cdot 2^{-12}\left(\frac{\tilde{\Lambda}_1}{a}\right)^6&{\rm for}&N_f=1\,,\\
\vspace{1ex}
      \displaystyle 
2^{-11}\left(\frac{\tilde{\Lambda}_2}{a}\right)^4&{\rm for}&N_f=2\,,\\
      \displaystyle 
2^{-10}\left(\frac{\tilde{\Lambda}_3}{a}\right)^2&{\rm for}& N_f=3\,.
    \end{array}\right.
\end{eqnarray}
The relation between the dynamical scales $\tilde{\Lambda}_{N_f}$ is given by,
\begin{eqnarray}
  \label{decouple}
  m^2 \tilde{\Lambda}_{N_f}^{8-2N_f}=\tilde{\Lambda}_{N_f-1}^{8-2 (N_f-1)}.
\end{eqnarray}
This decoupling relation also holds for $\Lambda_{N_f}$, which we have
examined in the instanton calculus of the massive N=2 supersymmetric QCD. 
Using the relation $\tilde{\Lambda}_0=\sqrt{2}\Lambda_0$, we obtain
the relation between the dynamical scales: $\tilde{\Lambda}^{8-2N_f}_{N_f}=
16\Lambda^{8-2N_f}_{N_f}$. From this relation, we find
that the microscopic instanton calculus
agrees\footnote{Note on the convention for $a$: The definition of
  $a$ differs between Eq.(\ref{instcal}) and
  Eq.(\ref{exactresult})  by 2.}
with the exact results
for $N_f=0,1,2$. However we also find a discrepancy between them for $N_f=3$.

In the similar way, we have evaluated the four-point function
$\VEV{\bar{\lambda}\bar{\lambda}\bar{\psi}\bar{\psi}}$ by the
  instanton calculus\cite{FP,DKM}, and have found that the non-trivial relation
  Eq.(\ref{u(a)}) holds for $N_f=0,1,2$.
For $N_f=3$, this four-point function does not depend on ${\cal F}_2$,
therefore it is not useful to check the exact result.

More detail and complete explanations of this letter will appear in
the near future\cite{AHSW}. 

\vspace{0.5cm}
\noindent
{\large \bf note added}  

\noindent
\begin{enumerate}
\item After the completion of this work, we learned that the four
  point function 
$\VEV{\bar{\lambda}\bar{\lambda}\bar{\psi}\bar{\psi}}$ was calculated
independently in \cite{DKM2}.
\item The formulas (\ref{vevu}) and
  (\ref{inf}) hold for $N_f=4$ by replacing $\Lambda_{N_f}^{8-2 N_f}$
  with $q=e^{-16\pi^2 /g^2 +2i\theta}$. 
In this case, 
\begin{eqnarray}
\VEV{u}_2=\frac{1}{2}a^{2}\times\frac{7}{2^5 3^5} q\quad.  
\nonumber
\end{eqnarray}
This result does not agree with the exact
result, which is based on the assumption that no quantum correction
appears in this case. After completing this calculation, we received the paper 
  \cite{DKM3}, in which the quantum correction in the $N_f=4$ theory was
discussed. 
We thank M.~P.~Mattis for informing us of the appearance of their preprint
and also thank K.~Ito and N.~Sasakura for the discussion on
this point.
\end{enumerate}

\vskip1cm
\centerline{\large\bf Acknowledgment}

\noindent
We thank our colleagues at Kyoto University for discussion
and encouragements.
We would especially like to acknowledge the numerous
valuable discussions with S.~Sugimoto.
The work of H.A. is  supported in part by the Grant-in-Aid
for Scientific Research (C)-07640391 and (C)-08240222 and 
those of T.H. and S.W. are supported in part by the Grant-in-Aid for
JSPS fellows.

\newcommand{\J}[4]{{\sl #1} {\bf #2} (19#3) #4}
\newcommand{\MPL}{Mod.~Phys.~Lett.}
\newcommand{\NP}{Nucl.~Phys.}
\newcommand{\PL}{Phys.~Lett.}
\newcommand{\PR}{Phys.~Rev.}
\newcommand{\PRL}{Phys.~Rev.~Lett.}
\newcommand{\AP}{Ann.~Phys.}
\newcommand{\CMP}{Commun.~Math.~Phys.}
\newcommand{\CQG}{Class.~Quant.~Grav.}
\newcommand{\PRP}{Phys.~Rept.}
\newcommand{\SPU}{Sov.~Phys.~Usp.}
\newcommand{\RMPA}{Rev.~Math.~Pur.~et~Appl.}
\newcommand{\SPJ}{Sov.~Phys.~JETP}
\newcommand{\MP}{Int.~Mod.~Phys.}
\newcommand{\ZP}{Z.~Phys.}

\end{document}